\documentclass[journal]{IEEEtran}
\IEEEoverridecommandlockouts

\usepackage{makecell}
\usepackage{array}
\usepackage{graphicx,amssymb,amsmath}
\usepackage{multicol}
\usepackage[noadjust]{cite}
\usepackage{setspace}
\usepackage{subfigure}
\usepackage{graphicx}
\usepackage{float}
\usepackage {url}
\usepackage{stfloats}
\usepackage{amsthm,pifont}
\usepackage{flushend}
\usepackage{cases,subeqnarray}
\usepackage{bm,multirow,bigstrut}
\usepackage{amsmath, amsthm, amssymb}
\usepackage{textcomp}
\usepackage{latexsym,bm}
\usepackage{booktabs}
\usepackage{xcolor}
\usepackage{mathtools}
\usepackage{dsfont}
\usepackage{extarrows}
\usepackage{epsfig}
\usepackage{epsfig}
\usepackage{epstopdf}
\usepackage[noend]{algpseudocode}
\usepackage{algorithmicx,algorithm}
\renewcommand{\algorithmicrequire}{\textbf{Input:}}
\renewcommand{\algorithmicensure}{\textbf{Output:}}

\usepackage{cite}
\usepackage{bm}
\usepackage{cleveref}
\usepackage{multicol}       
\usepackage{multirow}       
\usepackage{array}          
\usepackage{colortbl}
\usepackage{makecell}
\definecolor{crimson}{RGB}{192,0,0}         
\definecolor{navy}{RGB}{47,85,151}         
\usepackage{bbding}
\usepackage{graphicx}
\usepackage{booktabs}
\usepackage{algorithm}
\usepackage{algpseudocode}

\theoremstyle{plain}

\theoremstyle{plain}

\usepackage{amsmath}

\IEEEoverridecommandlockouts

\def\BibTeX{{\rm B\kern-.05em{\sc i\kern-.025em b}\kern-.08em
    T\kern-.1667em\lower.7ex\hbox{E}\kern-.125emX}}
\begin{document}

\title{Joint Power Allocation and Phase Shift Design for Stacked Intelligent Metasurfaces-aided Cell-Free Massive MIMO Systems with MARL}

\author{{Yiyang Zhu, Jiayi Zhang,~\IEEEmembership{Senior Member,~IEEE}, Enyu Shi, Ziheng Liu, \\Chau Yuen,~\IEEEmembership{Fellow,~IEEE}, and Bo Ai,~\IEEEmembership{Fellow,~IEEE}}
\thanks{Y. Zhu, J. Zhang, E. Shi, Z. Liu, and B. Ai are with the School of Electronic and Information Engineering, Beijing Jiaotong University, Beijing, China (e-mail: jiayizhang@bjtu.edu.cn).}
\thanks{C. Yuen is with the School of Electrical and Electronics Engineering, Nanyang Technological University, Singapore 639798, Singapore (e-mail: chau.yuen@ntu.edu.sg).}
}

\maketitle

\begin{abstract}
Cell-free (CF) massive multiple-input multiple-output (mMIMO) systems offer high spectral efficiency (SE) through multiple distributed access points (APs). However, the large number of antennas increases power consumption. We propose incorporating stacked intelligent metasurfaces (SIM) into CF mMIMO systems as a cost-effective, energy-efficient solution. This paper focuses on optimizing the joint power allocation of APs and the phase shift of SIMs to maximize the sum SE. To address this complex problem, we introduce a fully distributed multi-agent reinforcement learning (MARL) algorithm. Our novel algorithm, the noisy value method with a recurrent policy in multi-agent policy optimization (NVR-MAPPO), enhances performance by encouraging diverse exploration under centralized training and decentralized execution. Simulations demonstrate that NVR-MAPPO significantly improves sum SE and robustness across various scenarios.
\end{abstract}


\section{Introduction}
The forthcoming sixth generation (6G) network aims to meet the communication needs of both humans and intelligent machines, impacting future society, industry, and daily life \cite{zhang2020JSAC}. It integrates distributed networks with large-scale multiple-input multiple-output (MIMO) systems to enhance ultra-dense networks, known as cell-free (CF) massive MIMO (mMIMO) networks. In CF mMIMO networks, numerous access points (APs) are strategically deployed across extensive areas to mitigate shadow fading and shorten communication distances between transceivers \cite{nguyen2023twc}. These APs are connected to a central processing unit (CPU) via backhaul links and work collaboratively to serve all user equipment (UEs) simultaneously.

Despite performance improvements, concerns about energy consumption in CF mMIMO systems have increased with the deployment of numerous APs. Reconfigurable intelligent surfaces (RIS) have emerged as a promising solution to optimize system efficiency and reduce energy and costs \cite{10167480}. RIS is gaining traction for enhancing channel capacity, power efficiency, reliability, and wireless coverage \cite{wu2021tcomm,liu2021Star,dai2023twc,li2023tvt,shi2024ris}. However, current research focuses on single-layer metasurfaces, limiting beam pattern flexibility and multi-user interference suppression.



Inspired by wave-based computation, an all-optical diffractive deep neural network (D2NN) with multiple passive layers has been proposed \cite{Lin2018sci}. Initially constrained by its dependence on 3D-printed neurons, which restricted it to a single task, researchers have introduced a programmable D2NN that utilizes stacked intelligent metasurfaces (SIM). This advancement facilitates real-time adjustments of network coefficients through an intelligent controller \cite{Liu2022APD}. SIM technology maintains the rapid processing capabilities of the D2NN while enabling adaptability across various tasks \cite{an2023stacked}.

Advances in SIM technology could replace baseband beamforming in point-to-point communication systems. A novel hybrid digital-wave domain channel estimator initially processes received training symbols within the wave domain of the SIM layers before transitioning to the digital domain \cite{nadeem2023hybrid}. Additionally, a proposed analog beamforming scheme utilizing SIM significantly reduces precoding delay compared to digital methods \cite{an2023ICC}. However, challenges remain, particularly concerning the application of computationally intensive algorithms and joint learning in practical scenarios, necessitating further efforts to address deployment issues.

Machine learning is pivotal for the evolution of future wireless communication systems, particularly in addressing non-convex optimization problems. A distributed machine learning-based technique facilitates reliable downlink channel estimation \cite{dai2022dml}. A customized deep reinforcement learning approach enhances SIM-aided multi-user MIMO systems by integrating low-complexity transmit RF chains \cite{liu2024drlbased}. Additionally, a dynamic power control algorithm employing a novel multi-agent reinforcement learning (MARL) approach addresses high-dimensional signal processing challenges \cite{zhu2024marl}. Despite these advancements, developing effective constraint rules to improve agent cooperation and communication in SIM-aided systems remains an urgent challenge.



Building on these observations, we seek to improve the sum spectral efficiency (SE) by exploring the integration of joint power allocation and phase shifting in the SIM-aided CF mMIMO system. Considering the challenges of non-convex optimization, this paper introduces an innovative downlink design based on MARL to maximize sum SE. The primary contributions of this study are outlined as follows:
\begin{itemize}
    \item We explore a SIM-aided CF mMIMO system by formulating an optimization problem that integrates joint power allocation and phase shifting to maximize sum SE. Unlike the traditional MARL centralized training and execution (MARL-CTCE) approach, we employ centralized training with decentralized execution (MARL-CTDE), facilitating more effective cooperation and communication among MARL agents.
    \item Using MARL-CTDE, we design a two-layer network to manage AP power allocation and SIM phase shift optimization. Each AP-SIM combination requires partial global channel state information (CSI) and exchanges local observations, thus reducing the backhaul burden. Additionally, we introduce a noisy value method combined with a recurrent policy to enhance MARL exploration capabilities. Simulations show that the proposed algorithm outperforms existing MARL algorithms and significantly improves SE.
\end{itemize}

\textit{Notation}: The mathematical notation $(\cdot)^{H}$ denotes the conjugate transpose operation. Boldface uppercase letters such as $\mathbf{X}$ represent matrices, while boldface lowercase letters such as $\mathbf{x}$ denote column vectors. $\mathrm{sinc}(x) = \sin(\pi x) / (\pi x)$ is the $\mathrm{sinc}$ function. Furthermore, the complex Gaussian random variable $x$ with variance $\sigma^2$ is represented by $x \sim \mathcal{C}\mathcal{N}\left({0,{\sigma^2}} \right)$.


\section{System Model}

In this paper, we consider a SIM-aided CF mMIMO system with \textit{K} single-antenna UEs that are arbitrarily distributed in a large service area, as illustrated in Fig. 1. The UEs are jointly served by \textit{L} SIM-enhanced APs, each equipped with $M_{AP}$ antennas. All APs are connected to the CPU via fronthaul links which can send the AP data to the CPU for signal processing. Specifically, the SIM comprises \textit{M} metasurface layers, each containing \textit{N} meta-atoms. Additionally, the SIM is linked to a smart controller capable of applying an independent and adjustable phase shift to the electromagnetic (EM) waves transmitted through each meta-atom. By appropriately adjusting the phase shifts across the metasurfaces, the SIM performs downlink beamforming directly in the EM wave domain. Let $\mathcal{M}=\left\{1,2,\dots,M\right\}, \mathcal{N}=\left\{1,2,\dots,N\right\}, \mathcal{L} = \left\{1,2,\dots,L\right\}$ and $\mathcal{K}=\left\{1,2,\dots,K\right\}$ denote the index sets of SIM metasurface layers, meta-atoms per layer, APs, and UEs, respectively.

\subsection{Channel Model}
The resulting equivalent channel, denoted as $\hat{\mathbf{h}}_{l,k} \in \mathbb{C}^{N}$, originating from the \textit{l}-th SIM and targeting the \textit{k}-th UE, is expressed as
\begin{align}
    \hat{\mathbf{h}}_{l,k} = \beta_{l,k} \vert \mathbf{h}_{l,k} \vert ^{2},
\end{align}
where $\beta_{l,k}$ denotes the large-scale factor, $\mathbf{h}_{l,k}$ is the Rayleigh fading vector composed of the small-scale fading coefficients between AP \textit{l} and UE \textit{k}.

In our model, each layer of the SIM is represented as a uniform planar array arranged in a square configuration. For the $l$-th AP, let ${e^{j\varphi _{l,m}^n}},\forall n \in {\cal N},\forall m \in {\cal M}$ denote the phase shift imposed by the $n$-th meta-atom in the $m$-th meta-surface layer, where $\varphi _{l,m}^n \in \left[ {0,2\pi } \right)$ is the corresponding phase shift.Thus, the diagonal phase shift matrix for the $m$-th metasurface layer is denoted by ${{\bf{\Phi }}_{l,m}} = {\rm{diag}}\left( {e^{j\varphi _{l,m}^1}, e^{j\varphi _{l,m}^2}, \ldots, e^{j\varphi _{l,m}^N}} \right) \in \mathbb{C}^{N \times N}, \forall m \in \mathcal{M}$.

Furthermore, let ${{\bf{W}}_{l,m}} \in {\mathbb{C}^{N \times N}},\forall m \in \{ 1\} /{\cal M}$ be the transmission matrix from the $(m-1)$-th meta-surface layer to the $m$-th meta-surface layer and let ${{\bf{W}}_{l,1}} \in {\mathbb{C}^{N \times {M_{AP}}}}$ denote the transmission vector from the AP to the first meta-surface layer of the SIM. The propagation coefficient of the EM wave between adjacent metasurface layers can be derived from the Rayleigh-Sommerfeld diffraction equation. Thus, the $(n, n')$-th entry $w_{l,m}^{n, n'}$ of the propagation matrix ${{\bf{W}}_{l,m}}$ is denoted as

\begin{align}
    w_{l,m}^{n,n'} = \frac{{{d_x}{d_y}\cos \chi _{l,m}^{n,n'}}}{{d_{l,m}^{n,n'}}}\left( {\frac{1}{{2\pi d_{l,m}^{n,n'}}} - j\frac{1}{\lambda }} \right){e^{j2\pi \frac{{d_{l,m}^{n,n'}}}{\lambda }}},
\end{align}
for $\forall m \in \{ 1\} /{\cal M}$, where $\lambda$ is the wavelength, $d_{l,m}^{n,n'}$ denotes the transmission distance, $\chi _{l,m}^{n,n'}$ represents the angle between the propagation direction and the normal direction of the $(m-1)$-th meta-surface layer, and ${d_x} \times {d_y}$ is the size of each meta-atom. Similarly, the $n$-th entry $w_{l,m}^{n,1}$ of ${{\bf{W}}_{l,1}}$ is obtained from (2). Thus, the $l$-th SIM-enabled wave-based beamforming matrix $\mathbf{G}_{l} \in {\mathbb{C}^{N \times N}}$ is written as
\begin{align}
    {{\bf{G}}_l} = {{\bf{\Phi }}_{l,M}}{{\bf{W}}_{l,M}}{{\bf{\Phi }}_{l,M - 1}}{{\bf{W}}_{l,M - 1}} \ldots {{\bf{\Phi }}_{l,2}}{{\bf{W}}_{l,2}}{{\bf{\Phi }}_{l,1}}.
\end{align}

\begin{figure}[t]
    \centering
    \includegraphics[width=0.4\textwidth]{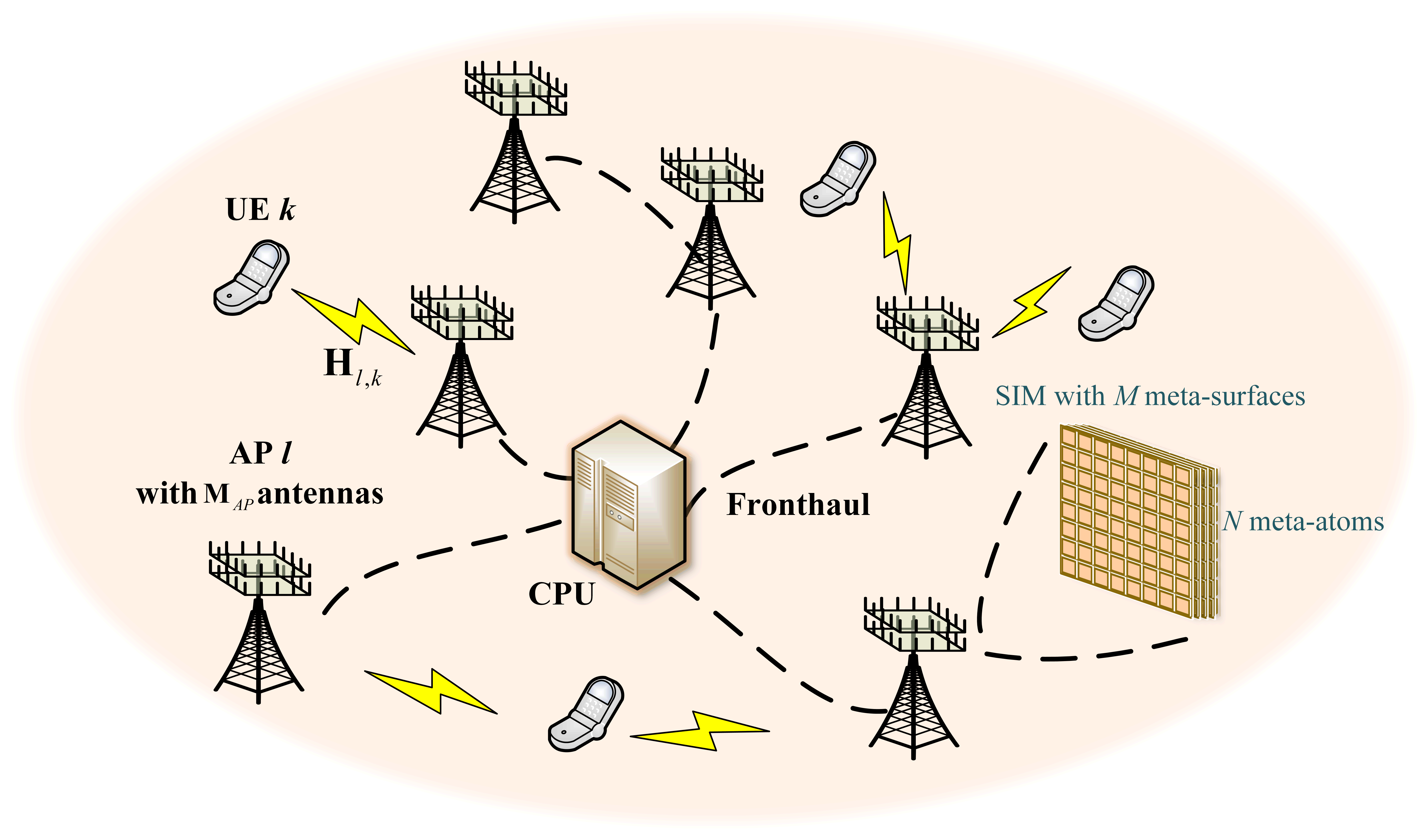}\vspace{-0.2cm}
    \caption{The SIM-aided CF mMIMO system.}
    \label{Fig1} \vspace{-0.3cm}
\end{figure}

\vspace{-0.2cm}
\subsection{Transmitters and Receivers}
Our proposed SIM-aided CF mMIMO system ensures synchronization among all APs, which is essential for enabling coherent joint transmission to serve all users. Let $\mathbf{s} \triangleq [s_{1}, s_{2}, \dots, s_{K}]^{T} \in \mathbb{C}^{K}$ denote the vector of symbols, where each $s_{k}$ represents the symbol transmitted to the \textit{k}-th user. Let $\mathbf{p}_{l,k} = \left[p_{l,k,1},\dots, p_{l,k,M_{AP}}\right]^T \in \mathbb{C}^{M_{AP}}, {p_{l,k,m}} \geq 0, \forall m \in \left[1,\dots, M_{AP}\right]$ denote the power allocated by the $l$-th AP to $k$-th UE. Let's represent the baseband frequency-domain signal $y_{k}^{}$ received by UE \textit{k} as 
\begin{align}
    {{y}_k} &= \mathop \sum \limits_{l = 1}^L {{\hat{\bf{h}}}_{l,k}^H{{\bf{G}}_l} {{\bf{W}}_{l,1}}\mathbf{p}_{l,k}{s_k}} + n_k,
\end{align}
where ${n_k}\sim{\cal C}{\cal N}(0,\sigma _k^2)$ denotes the i.i.d. additive white Gaussian noise and $\sigma_k^2$ is the noise power at the $k$-th UE.

\vspace{-0.2cm}
\subsection{Problem Formulation}
Based on the system model above, this subsection aims to enhance the overall SE gain realized over the network's operational duration.
At first, the signal-to-interference-and-noise ratio (SINR) for the transmitted symbol $s_{k}$ at UE \textit{k} is calculated as
\begin{equation}
\begin{aligned}
        {\gamma _k} = \frac{{|\mathop \sum \limits_{l = 1}^L {\hat{\bf{h}}}_{l,k}^H{{\bf{G}}_l}{{\bf{W}}_{l,1}}\mathbf{p}_{l,k}{|^2}}}{{\mathop \sum \limits_{j = 1,j \ne k}^K |\mathop \sum \limits_{l = 1}^L {\hat{\bf{h}}}_{l,k}^H{{\bf{G}}_l} {{\bf{W}}_{l,1}}{{\mathbf{p}_{l,j}}}{|^2} + \sigma _k^2}},\forall k \in {\cal K}.
\end{aligned}
\end{equation}

Thereby, the SE of UE \textit{k} $R_{k}^{}$ is given by
\begin{equation}
    \begin{aligned}
        R_{k}^{} = \text{log}_{2}(1+\gamma_{k}^{}).
    \end{aligned}
\end{equation}

Finally, the optimization problem of maximizing SE gain can be originally formulated as
\begin{equation}
\begin{aligned}
\mathcal{P}^{0} &\mathop {\max}\limits_{\mathbf{p}_{l,k},{\bf{\Phi}}_{l,m} } \text{sum SE} = \sum\limits_{k = 1}^K {{{\log }_2}\left( {1 + {\gamma _k}} \right)}, \\
&{\rm{s.t.}}\quad \mathop \sum \limits_{k = 1}^K ||{\mathbf{p}_{l,k}}||^{2} \le {P_{l,\max }},\quad \forall l \in {\cal L},\\ 
&\quad \quad \: {p_{l,k,m}} \ge 0, \quad\forall l \in {\cal L}, k \in {\cal{K}},\\
&\quad \quad \: \varphi _{l,m}^n \in [0,2\pi ),\quad\forall l \in {\cal L},\:\forall m \in {\cal M},\:\forall n \in {\cal N},
\end{aligned}
\end{equation}
where ${P_{l,\max }}$ is the maximum total power constraint of the $l$-th AP, $\varphi _{l,m}^n \in \left[ {0,2\pi } \right)$ is the corresponding phase shift. 

Given the complexity of the non-convex objective function (7), optimizing the power allocation and numerous phase shift matrices for each meta-surface layer is challenging. Inspired by MARL, we propose an innovative joint precoding network to address optimization problem $\mathcal{P}^{0}$, as detailed in Section III.

\section{Proposed Joint Power Allocation and Phase Shift Optimization Framework}

\subsection{Overview of the Framework}
\begin{figure*}[t]
    \centering
    \includegraphics[width=0.75\textwidth]{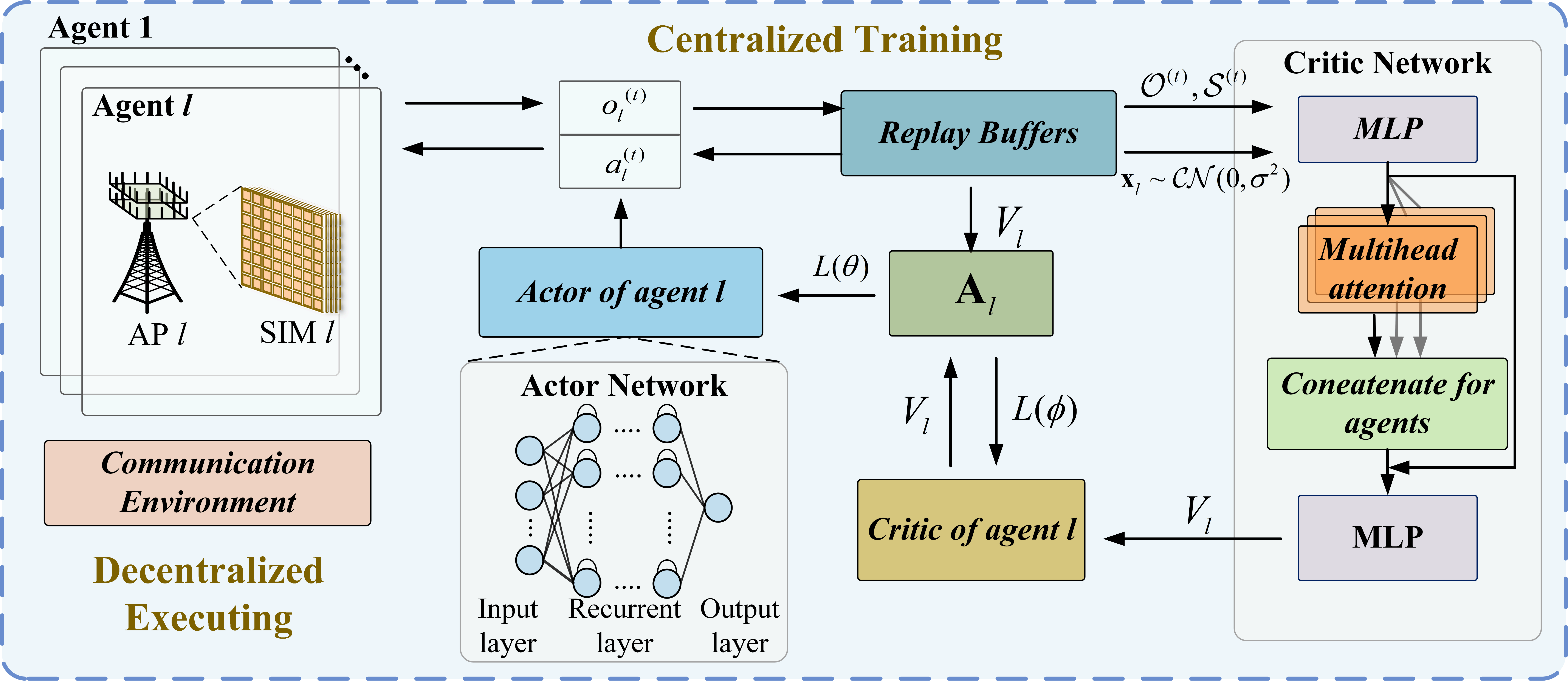}\vspace{-0.2cm}
    \caption{The proposed MARL precoding network.}
    \vspace{-0.3cm}
\end{figure*}

Under the framework of MARL-CTDE, illustrated in Fig. 2, a cooperative task is considered within a multi-agent environment. In this scenario, each agent comprises both an actor and a critic, which are responsible for action allocation and policy updates, respectively. The cooperative agents operate within a decentralized partially observable Markov decision process (Dec-POMDP) where rewards are shared.. Sequential actions are chosen by the agents under conditions of partial observation and environmental stochasticity. Hence, we consider each AP-SIM group as the agent and describe the proposed downlink optimization problem with an MARL tuple $<\mathcal{S}^{(t)},\mathcal{O}^{(t)},\mathcal{A}^{(t)},\mathcal{P}^{(t)},\mathcal{R}^{(t)}>$ at slot \textit{t}, where $\mathcal{S}^{(t)}$, $\mathcal{O}^{(t)}$, $\mathcal{A}^{(t)}$,$\mathcal{P}^{(t)} (\mathbf{s}'|\mathbf{s},\mathbf{A}^{(t)})$ and $\mathcal{R}^{(t)}$ denotes the state space, the local observation space, the joint action space, and the transition probability from $\mathbf{S}$ to $\mathbf{S}'$ given the joint action $\mathbf{A}^{(t)}$ for all $L$ agents, and shared rewards at slot \textit{t}, respectively.

\vspace{-0.2cm}
\subsection{Multi-Agent Proximal Policy Optimization}
\subsubsection{Agent-Specific State and Observation}
While local observations in the CF mMIMO environment do not include information about channels from other agents to specific UEs, they provide agent-specific features such as the agent ID, available actions, SIM information, and local CSI, which are absent from the global state. Therefore, we employ an observation method that integrates both global and partial state variables, where each agent shares its local observation with the central critic network. For time slot $t$ and AP-SIM group $l$, the global state and local observation information are considered as follows:
\begin{align}
    \mathcal{S}_{l}^{(t)} \!\!\!=\!\! \left(\textbf{\textit{D}},\mathbf{W}_{l,m},\gamma_{k}^{(t)} \right),\quad
    \mathcal{O}_{l}^{(t)} \!\!\!=\!\! \left(\mathbf{H}_{l,k}^{(t)},\mathbf{p}^{(t)}_l,\!\mathbf{\Phi}_{l}^{(t)} \right)\!, 
\end{align}
where $\textbf{\textit{D}}$ is the relative positions of all UEs and APs, $\mathbf{p^{(t)}}_{l} = [p_{l,1},p_{l,2},\dots,p_{l,K}]$ is the power vector of AP $l$, and $\mathbf{\Phi}^{(t)}_{l}$ if the phase shift of the $l$-th SIM at slot \textit{t}, respectively.

\subsubsection{Actor and Critic Network}
In the proposed framework, we utilize two separate networks: an actor network with parameters $\theta$ and a value function network with parameters $\phi$. Although each agent could potentially have its own pair of actor and critic networks, for simplicity and without loss of generality, we assume that all agents share the same critic and actor networks. Specifically, in a MARL scenario, we expand PPO-clip to optimize the independent policies of the agents,
\begin{equation}
\begin{aligned}
    J&^{\text {MAPPO-clip }} = \\
    &\frac{1}{N} \sum_{i}^{N} \mathbb{E}\left[\min \left(r^{i} \mathbf{A}^{\text {old }}, \operatorname{clip}\left(r^{i}, 1\!-\!\epsilon, 1\!+\!\epsilon\right) \mathbf{A}^{\text {old }}\right)\right],
\end{aligned}
\end{equation}
where $\pi$ is the policy, $r^{i}=\frac{\pi^{i}\left(a^{i} \mid \tau^{i}\right)}{\pi_{\text {old }}^{i}\left(a^{i} \mid \tau^{i}\right)}$ is the importance sampling weight, which approximates the KL-divergence constraint.
\subsection{Recurrent-MAPPO with Noisy Values for Maximizing sum SE of SIM-aided CF MIMO}

The high-dimensional phase shift matrix in SIM complicates gradient estimation, requiring many samples to characterize the state effectively. This issue is worsened by the dense presence of UEs, APs, and SIMs. Additionally, the approximate centralized value function has high bias at the start of training. To address these issues, we propose a policy regularization method, noisy values (NV), and integrate a recurrent policy into MAPPO, named NVR-MAPPO.
\addtolength{\topmargin}{+0.05 in}
\begin{algorithm}[!t]
    \caption{NVR-MAPPO Algorithm for Maximizing sum SE}
    \label{alg:HATRPO}
    \renewcommand{\algorithmicrequire}{\textbf{Input:}}
    \renewcommand{\algorithmicensure}{\textbf{Output:}}
    \begin{algorithmic}[1]
    \State \textbf{Input} Batch size $\mathcal{B}$,  number of: agents \textit{L}, episodes $\hat{N}$, steps per episode \textit{T},  noise weight $\alpha$, entropy loss weight $\eta$ and state space $\mathcal{S}$
    \State Sample Gaussian noise $\textbf{x}_{l}\sim \mathcal{CN}(0,1), \forall l \in \mathcal{L}$
    \For{ $\hat{n}$ = 0, 1, $\ldots$, $\hat{N}$-1}
        \For{$t$ = 1, 2, $\ldots$, $T$}
            \State Involve the collection of a set of trajectories 
            \State Execute actions $\vec{a}^{(t)}$, observe $r^{(t)}, s^{(t+1)}, o^{(t+1)}$
            \State Push MARL tuple into $\mathcal{B}$
        \EndFor
        \State \textbf{if} \textit{NV shuffle interval} \textbf{then}
        \State Shuffe the noise in agent dimension
        \State \textbf{end if}
        \State Randomly select a minibatch from $\mathcal{B}$
        \State Compute noise value
        \For {agent $l_{\hat{l}}=l_{1}, \ldots, l_{n}$}
            \For {\textit{each data chunk} in the minibatch}
                \State update RNN hidden states policy
            \EndFor
            \State Compute the gradient of the max sum SE
            \State Update critic by minimizing the loss $L(\phi)$
            \State Update agent policy by using loss $L(\theta)$
        \EndFor
    \EndFor
    \end{algorithmic}

\end{algorithm}

For each agent $l$, we generate a Gaussian noise vector $\textbf{x}_{l}$ by sampling from $\mathcal{CN}(0, \sigma^2)$, where $\sigma^2$ denotes the noise variance. Subsequently, we concatenate the noise vector $\textbf{x}_{l}$ with the global state $\mathcal{S}$ and feed the concatenated features $\left(\mathcal{S}, \textbf{x}_{l}\right)$ into the centralized value network, which produces a value $v{l}^{(t)}$ for each agent. Correspondingly, the critic network is trained to minimize the mean-squared Bellman error function.
\begin{align}
    L(\phi)=\frac{1}{B \cdot L} \sum_{b=1}^{B} \sum_{l=1}^{L}\left(v_{b,l}^{(t)}(\phi)-\hat{R}_{b,l}^{(t)}\right)^{2},
\end{align}
where $B$, $\hat{R}_{b,i}^{(t)}$ denote the batch size and reward with noisy values, respectively. The actor network is trained to maximize
    \begin{align}
           &L(\theta ) = \frac{1}{{B \cdot L}}\sum\limits_{b = 1}^B {\sum\limits_{l = 1}^L {\left[ {\min \left( {r_b^{(t)}(\theta ),} \right.} \right.} } \notag\\
           &\left. {\left. {{\mathop{\rm clip}\nolimits} \left( {r_b^{(t)}(\theta )\hat A_b^{(t)},1 - ,1 + } \right)\hat A_b^i} \right) - \eta {\cal H}\left( {\pi _\theta ^{(t)}\left( {o_b^{(t)}} \right)} \right)} \right],
    \end{align}
where $r_{b}^{i}(\theta)=\frac{\pi_{\theta}^{i}\left(a_{b}^{i} \mid o_{b}^{i}\right)}{\pi_{\theta_{o l d}}^{i}\left(a_{b}^{i} \mid o_{b}^{i}\right)}, \forall i \in N, b \in B$,$\hat{A}_{b}^{(t)}$ is calculated by GAE method, $\eta$ is the entropy weight loss and $\mathcal{H}$ is the Shannon Entropy, respectively.

Finally, to ensure that the target network tends to be stable in the iterative process, we adopt recurrent policy in which the loss functions additionally sum over time.The procedure of the NVR-MAPPO for maximizing SE performance is summarized in Algorithm 1.

\section{Simulation Results}
In the proposed SIM-aided CF mMIMO network simulation, we consider a 100m $\times$ 100m area with non-cellular network services, where all UEs are served by all APs simultaneously. Each AP, integrated with a SIM for transmit beamforming, divides the area into $L$ squares with randomly positioned single-antenna UEs. The AP height is 10m, UE height is 1.7m, and SIM thickness is $T_{SIM} = 5\lambda$ at a 28 GHz downlink frequency ($\lambda = 10.8$ mm), making the spacing between adjacent metasurfaces in an $M$-layer SIM $d_{layer} = T_{SIM}/M$. All metasurfaces have $N \times N$ meta-atoms with half-wavelength spacing between adjacent antennas/meta-atoms, where each meta-atom size is $d_{x} = d_{y} = \lambda/2$. The simulation parameters include a maximum transmit power of $P_{l, \text{max}} = 3$ dBm per AP, an initial $M_{AP} = 4$ antennas per AP, and a noise power of $\sigma^{2} = -96$ dBm. Given the limited antennas and low transmit power in CF mMIMO networks, we use the channel model from \cite{zhang2021tsp}.

\subsection{Convergence of the proposed algorithm}
\begin{figure}[t]
    \centering
    \includegraphics[width=0.4\textwidth]{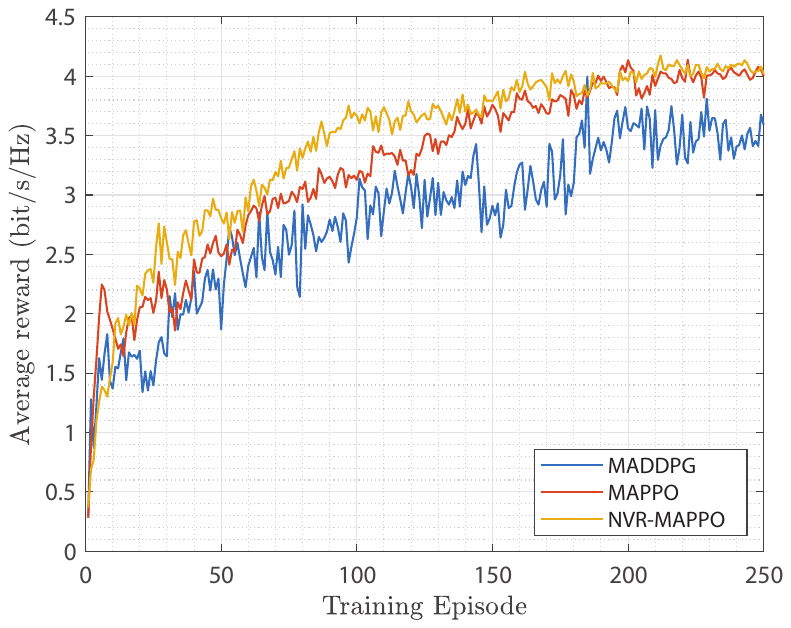}
    \caption{Average reward against the training episode with $episode$ = 250, $L$ = 8, $K$ = 4, $M_{AP}$ = 2, $M$ = 4, and $N$ = 64.}
    \label{Fig2} \vspace{-0.2cm}
\end{figure}

\begin{figure}[t]
    \centering
    \includegraphics[width=0.4\textwidth]{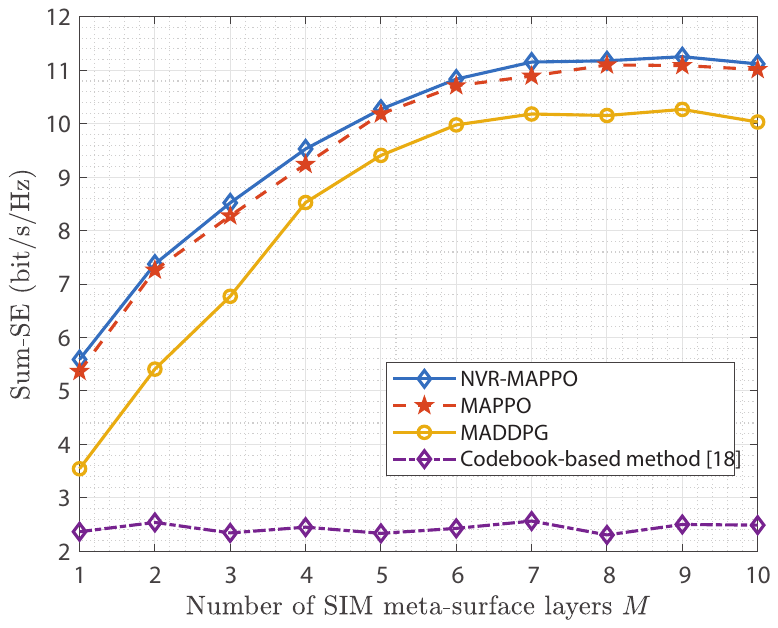}\vspace{-0.2cm}
    \caption{sum SE against the number of SIM meta-surface layers with $L$ = 8, $K$ = 6, $M_{AP}$ = 2, and $N$ = 32.}
    \label{Fig3} \vspace{-0.3cm}
\end{figure}

\begin{figure}[t]
    \centering
    \includegraphics[width=0.4\textwidth]{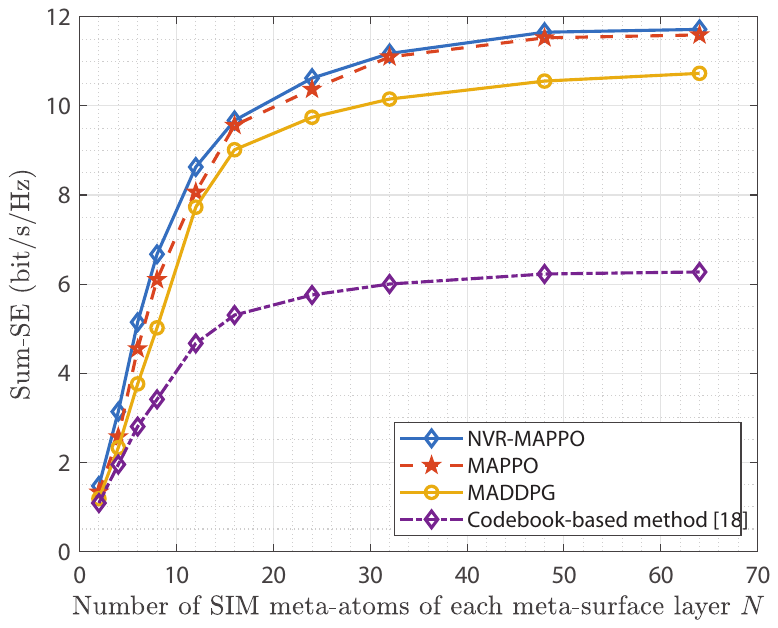}\vspace{-0.2cm}
    \caption{sum SE against the number of SIM meta-atoms of each SIM meta-surface layer with $L$ = 8, $K$ = 6, $M_{AP}$ = 2, and $M$ = 8.}
    \label{Fig4} \vspace{-0.3cm}
\end{figure}

To demonstrate algorithm convergence, we present a plot of reward versus training episodes in Fig. 3. The results indicate that MAPPO converges more quickly than MADDPG over the same number of training steps. Although the final convergence levels are similar, NVR-MAPPO achieves faster initial convergence compared to traditional MAPPO. The increased fluctuation observed with NVR-MAPPO is attributed to noise, which promotes exploration and accelerates convergence. Overall, MAPPO exhibits more stable convergence than MADDPG, suggesting superior practical stability.

\subsection{Impact of key system parameters} 
We evaluate the sum SE of the proposed SIM-aided cell-free network in this subsection.

\subsubsection{SE against the number of SIM meta-surface layers} 
Fig. 4 illustrates the sum spectral efficiency (SE) relative to the number of SIM meta-surface layers. As the number of meta-surface layers increases, the sum SE also rises. During the same training period, MARL methods enhance performance by at least 50\% compared to traditional codebook-based methods, which utilize random phase shift vectors and iterative water-filling power allocation \cite{an2022codebook}. NVR-MAPPO and MAPPO each demonstrate a performance improvement of at least 15\% over MADDPG. Although NVR-MAPPO provides only a marginal advantage over MAPPO, its faster training speed makes it more suitable for a broader range of continuous scenarios.

\subsubsection{SE against the number of SIM meta-atoms of each SIM meta-surface layer} 
MARL methods outperform traditional codebook approaches by approximately 50\%. Initially, with fewer meta-atoms, NVR-MAPPO and MAPPO do not exhibit significant advantages over MADDPG. However, as the number of meta-atoms increases, NVR-MAPPO and MAPPO show substantial improvements, achieving a 9.8\% enhancement at $N = 64$. The faster convergence of NVR-MAPPO underscores its robustness and suitability for a broader range of applications, approaching optimal performance.

\section{Conclusions}\label{se:conclusion} In this paper, we explored the maximization of downlink sum SE in a SIM-aided CF mMIMO system through power allocation and phase shift design. We proposed a MARL-based approach that integrates noisy value methods with a recurrent policy to foster diverse exploration and reduce computational time, making it well-suited for large networks. Our simulations validated the effectiveness of SIM in alleviating multi-user interference, demonstrating that the NVR-MAPPO method provides approximately an 18\% convergence advantage over MAPPO and offers improved reliability compared to traditional codebook-based algorithms within limited training time.

\bibliographystyle{IEEEtran} \bibliography{IEEEabrv,Bibliography}
\end{document}